\documentclass[aps, prl,twocolumn, superscriptaddress,amsmath, amssymb]{revtex4} 
\usepackage{graphicx}
\usepackage{amsmath}
\usepackage{dcolumn}
\usepackage{bm}
\usepackage{lmodern}
\usepackage[T1]{fontenc}
\usepackage{esint}

\usepackage[colorlinks,bookmarks=false,citecolor=blue,linkcolor=blue,urlcolor=blue]{hyperref}

\begin{document}

\title{Delocalization of vortex magnetic field near a planar defect}

\author{V. Plastovets} 
\affiliation{University of Bordeaux, LOMA UMR-CNRS 5798, F-33405 Talence Cedex, France}
\author{A. Buzdin}
\affiliation{University of Bordeaux, LOMA UMR-CNRS 5798, F-33405 Talence Cedex, France}
\affiliation{World-Class Research Center “Digital Biodesign and Personalized Healthcare”, Sechenov First Moscow State Medical University, 19991 Moscow, Russia}
\date{\today}

\begin{abstract}
In this letter we discuss physical mechanisms for delocalization of the magnetic field of Abrikosov vortex in superconductors in the presence of planar defects of various electronic transparency. The resulting strong perturbation of the supercurrent can significantly affect the local flux measurements and mimic the vortices that carried only part of a flux quantum. 
\end{abstract}

\maketitle

Very recently in Ref. \cite{FracVort} the observation of the superconducting vortices carrying a flux smaller than the flux quantum $\Phi_{0}=\pi\hbar c/e$ in the Ba$_{1-x}$K$_{x}$Fe$_{2}$As$_{2}$ superconductor was reported. These fractional vortices were observed in the hole overdoped Ba$_{1-x}$K$_{x}$Fe$_{2}$As$_{2}$ (${x=0.77}$) compound with a critical temperature $T_{c}\sim11$K which is substantially smaller the critical temperature of $38$K for the optimally doped $x=0.4$ compound \cite{Optimaldoped}. The fractional vortices coexist with the standard single quantum vortices but disappear below $9$K. In Refs. \cite{FracVort,Babaev} it was suggested that the appearance of the fractional vortices is related with the multiband superconductivity in Ba$_{1-x}$K$_{x}$Fe$_{2}$As$_{2}$ when the vortex exists only in one band.

This interpretation arises some doubts because of the inevitable interband interaction in 3D Ba$_{1-x}$K$_{x}$Fe$_{2}$As$_{2}$ superconductor which near $T_{c}$ effectively results in a single order parameter superconductivity with usual vortices. In present article we analyze how the vortex magnetic field distribution can be modified by the presence of the nearby planar defects of different type and demonstrate that vortex itself can carry a flux substantially smaller than $\Phi_{0}$, while the remaining flux may be strongly delocalized at distances of the order of the defect's size. Similar situation may be realized if the Abrikosov vortex is situated near the Josephson junction \cite{Fistul,Mironov} - it also can carry only a part the flux quantum while the remaining flux will be absorbed by the part of the Josephson vortex. As the result a hybrid Arikosov-Josephson vortex satisfies the condition of flux quantization but the flux of Abrikosov vortex (located at the distance of the order of London penetration depth $\lambda$) is smaller than $\Phi_{0}$ and the remaining flux is accumulated at the distances of the order of Josephson length $\lambda_{J}\gg\lambda.$ We believe that on experiment this situation may be seen as a fractional "bright" Abrikosov vortex supplemented by remaining weak field existing at distances much larger than $\lambda$ and hardly detected above the noise level. Note that different types of defects can greatly alter the superconducting current distribution, which can even mimic a generation of vortex-antivortex pairs \cite{Ge, Mironov2}.

We consider a planar defect of the thickness $d\ll \lambda$ lying in the plane $y=0$ whose transparency is characterized by angle-averaged probability of electron transmission $\mathcal{T}$, which determines the amplitude of the critical current density $j_c$ and is temperature independent \cite{Golubov}. 
In real crystals, the role of such defects can be played by the grain boundaries or twinning planes, which are often present in superconductors, can reach macroscopic sizes, and exhibit strong interaction with vortices (for example in Sn, Nb \cite{Khlyustikov}, cuprates \cite{Herbsommer} or pnictides \cite{Kalisky}).
The previous experiments on vortex imaging by magnetic force microscopy in optimally doped Ba$_{0.6}$K$_{0.4}$Fe$_{2}$As$_2$ and underdoped Ba$_{0.77}$K$_{0.23}$Fe$_{2}$As$_{2}$ single crystals \cite{Yang} clearly shows the presence of vortex chains near the twin boundaries serving as the strong pinning centers. The vortex interaction with a twin boundary may be either repulsive or attractive. If the superconductivity is locally enhanced near the boundary vortices should be repealed by it. Apparently such a situation is realized in underdoped BaFe$_2$(As$_{1-x}$P$_x$)$_2$ \cite{Yagil} and underdoped Ba(Fe$_{1-x}$Co$_{x}$)$_2$As$_2$ \cite{Kalisky}. On the other hand, the low transparency of the twin boundary should generate vortex attraction, which was observed in YBa$_2$Cu$_3$O$_{7-\delta}$ \cite{Herbsommer}.

\begin{figure}[] 
\begin{minipage}[h]{1.0\linewidth}
\includegraphics[width=1.0\textwidth]{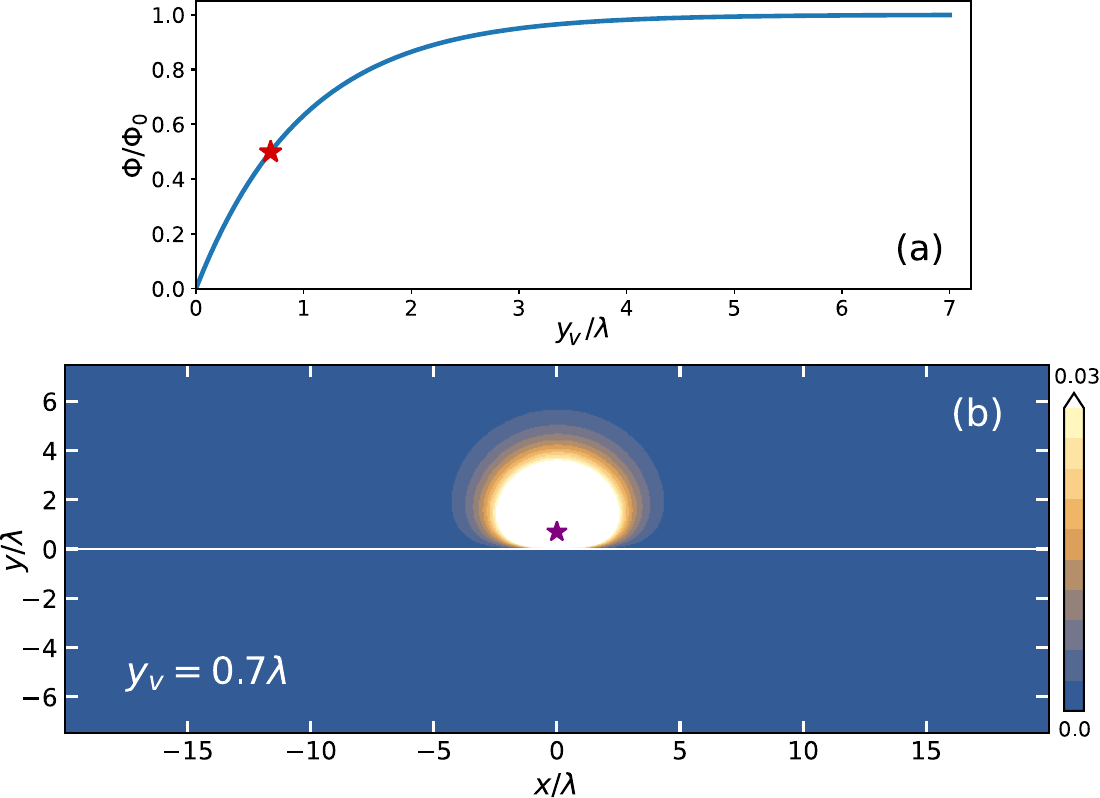} 
\end{minipage}
\caption{\small{ (a) Dependence of the magnetic flux on the vortex distance $y_v$ from the infinite isolating plane. (b) Spatial distribution of the magnetic field $h_z/\frac{\Phi_0}{2\pi \lambda^2}$ of the vortex located at $y_v=0.7\lambda$ (marked by the star) from the plane (shown by the white line). }}
\label{fig1}
\end{figure}

\

\textit{Vortex near a long planar defect with $\mathcal{T}=0$}.
Strongly insulating plane is nontransparent for the supercurrent, hence the Abrikosov vortex placed nearby has the same magnetic field distribution as a vortex near the edge of the sample \cite{De Gennes}. The latter is defined by a solution of the London equation with corresponding boundary conditions of vanishing supercurrent through the defect's plane 
\begin{gather}\label{LNDN}
    (\nabla^2-\lambda^{-2})h_z=\frac{\Phi_0}{2\pi\lambda^2}\delta({\bf r}_v); \
     \big(\nabla\times {\bf h} \cdot {\bf y}_0\big)|_{x\in(-\infty,\infty)}=0, 
\end{gather}
where $\delta({\bf r})$ is delta function.
Using the image method \cite{De Gennes} the magnetic field can be written as a superposition of the fields of a free vortex placed at the point $(x_v, \ y_v>0)$
$$
{\bf h}_v({\bf r}) = \frac{\Phi_0}{2\pi \lambda^2}K_0\Big[ \frac{|{\bf r}-{\bf r}_v|}{\lambda} \Big]{\bf z}_0
$$
and mirrored antivortex, such that
\begin{gather}\label{h_long}
h_z({\bf r}) = 
\begin{cases}
h_v({\bf r})-h_v({\bf r})|_{y_v\rightarrow -y_v}, \ y \geqslant  0 \\
0, \ y<0
\end{cases}.
\end{gather}
The total magnetic flux of the vortex may be easily calculated and is given by the expression \cite{Mironov}
\begin{gather}\label{Eq1}
\Phi\left(  y_v\right)  =\Phi_{0}\left[  1-\frac{y_v}{\pi\lambda
}\int_{-\infty}^{\infty}\frac{K_{1}\left(\frac{y_v\sqrt{1+t^{2}}}{\lambda
}\right)}{\sqrt{1+t^{2}}}  dt\right],
\end{gather}
where $y_v$ is the distance between vortex and the boundary and $K_i$ is modified Bessel function. At $y_v\gg\lambda$ the flux is equal to its standard value $\Phi_{0}$, but it vanishes when ${y_v\rightarrow0}$. The evolution of the vortex flux $\Phi\left(\frac{y_v}{\lambda}\right)$ is presented in Fig. \ref{fig1}(a). Naturally, a similar situation occurs in a superconductor with planar defects provoking a local weakening of superconductivity.

\begin{figure}[] 
\begin{minipage}[h]{1.0\linewidth}
\includegraphics[width=0.9\textwidth]{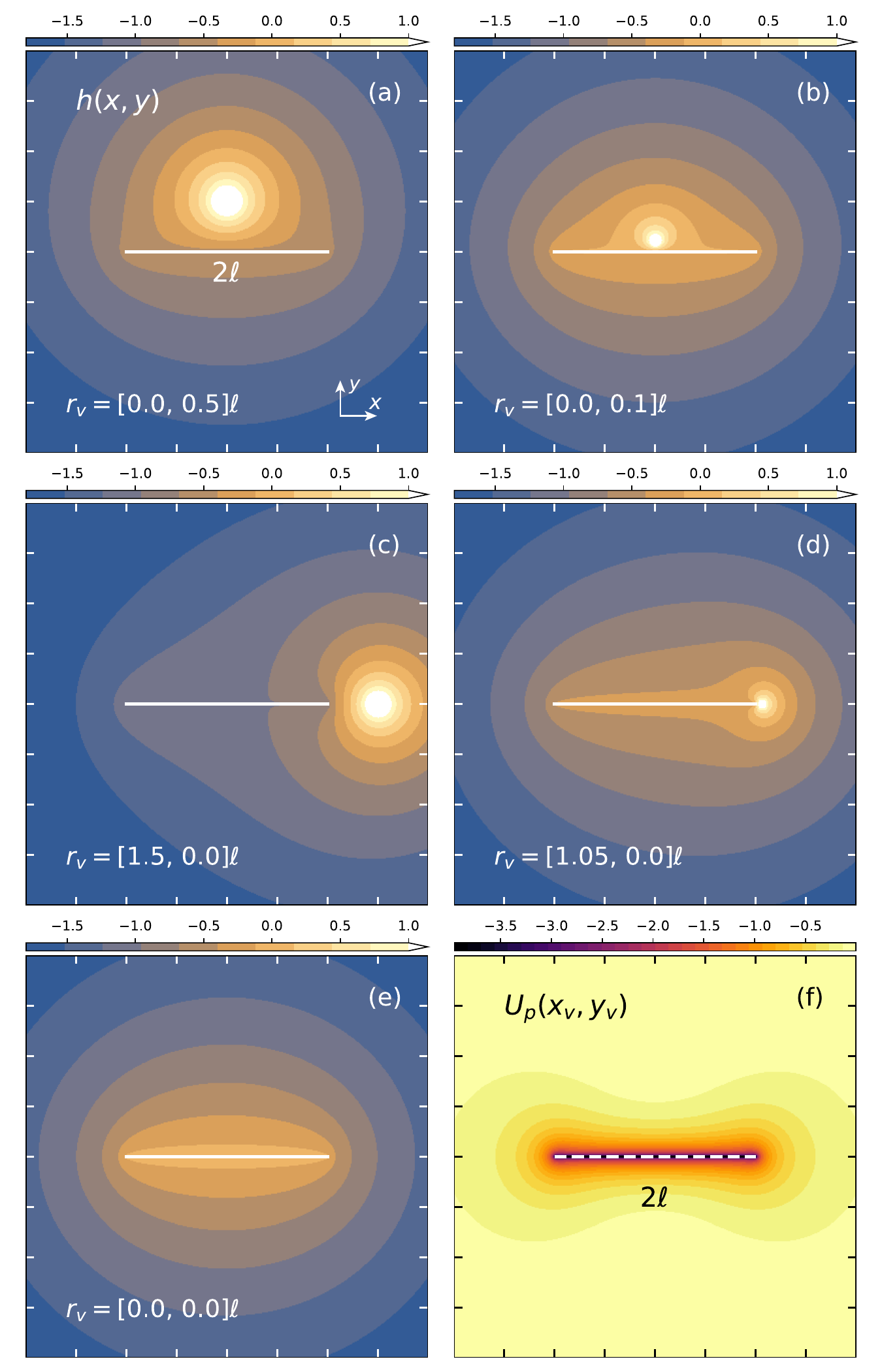} 
\end{minipage}
\caption{\small{ (a-e) Distribution of the magnetic field $h_z(x,y)/\frac{\Phi_0}{2\pi \lambda^2}$ of the Abrikosov vortex in the vicinity of the isolating plane (shown by the white line) of the length $2\ell$ for different vortex position $r_v=[x_v, y_v]\ell$. (f) Pinning energy profile $U_p/(\frac{\Phi_0}{2\pi \lambda^2})^2$ as a function of the vortex position ${\bf r}_v/\ell$.  }}
\label{fig2}
\end{figure}

\textit{Vortex near a short planar defect with $\mathcal{T}=0$}.
Here we consider an Abrikosov vortex near finite-sized insulating $xz$-plane of the size $2\ell$ in the $x$-direction. The distribution of the vortex magnetic field can be easily obtained from Eq. (\ref{LNDN}) with the same boundary conditions, which at short distances $|{\bf r}|\ll \lambda$ can be approximated as a Poisson equation \cite{brison}
\begin{gather}
    \nabla^2h_z=\frac{\Phi_0}{2\pi\lambda^2}\delta({\bf r}_v); \quad \big(\nabla\times {\bf h} \cdot {\bf y}_0\big)|_{x\in(-\ell,\ell)}=0. 
\end{gather}
This problem is similar to the electrostatic one and can be solved by conformal transformation method \cite{Buzdin}. The solution reads
\begin{gather}\label{Eq2}
    h_z({\bf r})=\frac{\Phi_0}{2\pi\lambda^2}\Bigg[\ln\frac{\lambda}{2\ell}+ \ln\Big| \frac{2}{e^{\gamma+i\varphi}- e^{\gamma_v+i\varphi_v}} \Big| 
    \\ \notag
    + \ln\Big| 1 -  e^{-(\gamma+\gamma_v)-i(\varphi-\varphi_v)}  \Big| \Bigg],
\end{gather}
where
\begin{gather}\notag
\gamma({\bf r})=\frac{1}{2}\cosh^{-1}\Bigg[ \frac{r^2}{\ell^2}+\sqrt{\Big( \frac{r^2}{\ell^2}-1 \Big)^2+4\frac{y^2}{\ell^2}} \Bigg],
\end{gather}
the phase is defined as ${e^{i\varphi(x,y)}= x \ \text{sech}(\gamma)/\ell+iy \ \text{csch}(\gamma)/\ell}$;
and index $v$ means vortex coordinates. 
Note that at $\ell\ll\lambda$ the magnetic field of the vortex changes slightly, while at $\ell\gg\lambda$ the strong deformation is described by the limiting case (\ref{h_long}). The most nontrivial redistribution of the field occurs in the case ${\ell\approx\lambda}$, and the corresponding illustrative examples are presented in Fig. \ref{fig2}. We clearly see that substantial part of vortex flux comes from the remote region with a weak magnetic field, especially when the vortex is placed close to the plane. When the vortex is pinned by the plane its magnetic field is smeared over the entire defect and the flux is strongly delocalized [Fig. \ref{fig2}(e)].

Interaction with a defect is described by the potential energy which can be calculated using the field (\ref{Eq2}) and reads up to a constant \cite{Bespalov}
\begin{gather}\label{U_p}
U_p = \Big(\frac{\Phi_0}{4\pi\lambda}\Big)^2 \Bigg[ \ln\Big( 1- e^{-2\gamma_v}   \Big) 
+ 
\ln\Big|  1 -  e^{-2(\gamma_v+i\varphi_v)}   \Big| \Bigg].
\end{gather}
The interaction of the plane segment and the vortex line is purely attractive and has anisotropy as shown in Fig. \ref{fig2}(f). The pinning force ${\bf F}_p$ in the vicinity of the defect (for $\ell\approx\lambda$) can be estimated in the limit ${\delta x_v, \delta y_v \ll \ell}$ as follows
\begin{gather}\notag
F_{px} (\ell+\delta x_v, 0) \approx \Big[ -\frac{1}{\delta x_v}+\sqrt{\frac2\ell}\frac{1}{\sqrt{\delta x_v}} \Big]\Big(\frac{\Phi_0}{4\pi\lambda}\Big)^2, \\ \notag
F_{py} (0, \delta y_v) \approx -\frac{1}{\delta y_v} \Big(\frac{\Phi_0}{4\pi\lambda}\Big)^2
\end{gather}
and we see that the vortex attraction is much stronger near the central part of the defect comparing with its edges. 
Conversely, away from the defect (for the case $\ell\ll\lambda$) the force scales as 
\begin{gather}\notag
F_{px} (x_v\gg\ell, 0) \approx -\frac{\ell^2}{x_v^3}\Big(\frac{\Phi_0}{4\pi\lambda}\Big)^2, \\ \notag
F_{py} (0, y_v\gg\ell) \approx -\frac{\ell^4}{4y_v^5}\Big(\frac{\Phi_0}{4\pi\lambda}\Big)^2
\end{gather}
and attraction from the edges becomes dominant.

\textit{Vortex near the plane with $\mathcal{T}\ll1$}.
Now we consider the case of the Abrikosov vortex situated near a long
Josephson junction with $\ell \gtrsim \lambda_J$ which can model a situation when the vortex is near a
segment-like defect with low transparency (or small critical current).
The Abrikosov vortex creates a local phase difference at the Josephson
junction at the nearby region $\sim\lambda$, which at the scale of the
Josephson length $\lambda_{J}\gg\lambda$ can be considered as a local phase
jump. This phase jump $\kappa$ is generated by vortex current at the
junction and is directly related with the vortex flux $\Phi$ from Eq. (\ref{Eq1}).
\begin{gather}
\kappa=2\pi\frac{\Phi_{0}-\Phi(y_v)  }{\Phi_{0}}.%
\end{gather}
\begin{figure}[] 
\begin{minipage}[h]{1.0\linewidth}
\includegraphics[width=0.95\textwidth]{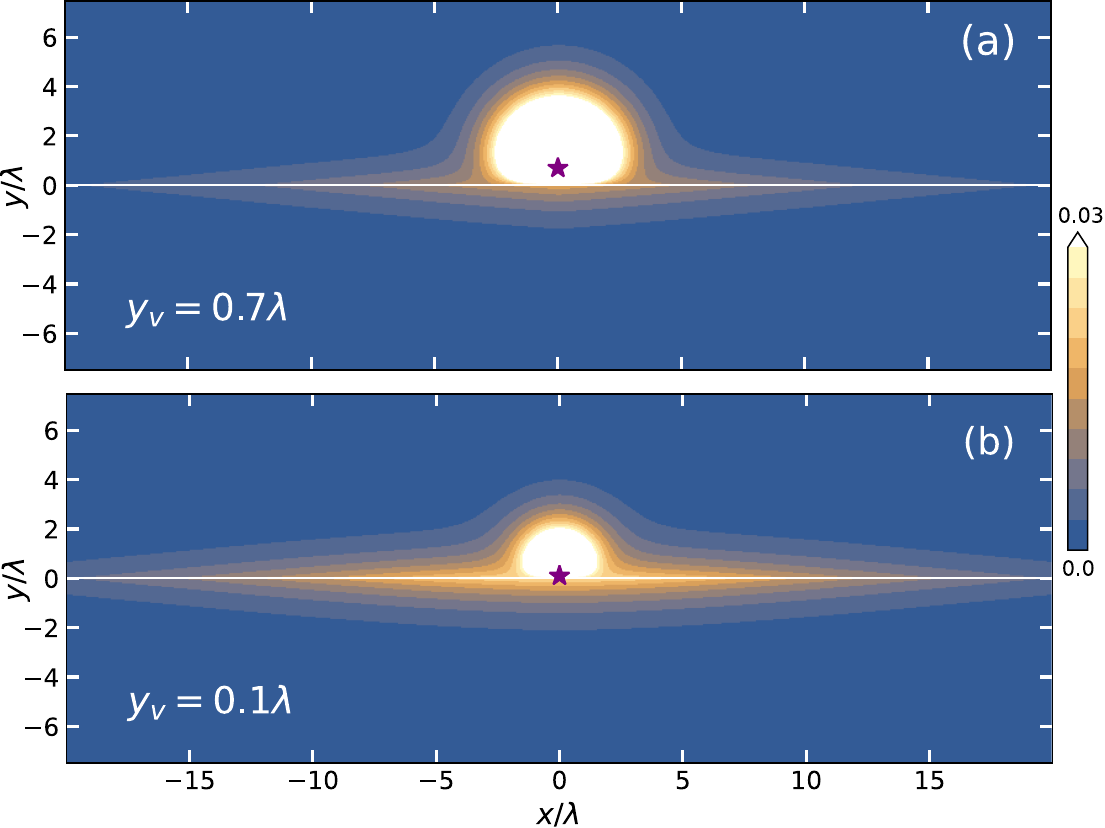} 
\end{minipage}
\caption{\small{ Spatial distribution of the magnetic field $h_z/\frac{\Phi_0}{2\pi \lambda^2}$ of the Abrikosov vortex located at $y_v$ (marked by the star) in the vicinity of the infinite Josephson junction (shown by the white line) for $\lambda_J/\lambda=10$. }}
\label{fig3}
\end{figure}

This situation is somewhat similar to the Josephson junction with a pair of
tiny current injectors generating a local phase jump \cite{Gold}. In our case a vortex plays a role of these injectors. The evolution of the phase
difference $\varphi$ on the Josephson junction is described by the
Ferrell-Prange equation \cite{Barone}%
\begin{gather}
\frac{d^{2}\varphi}{dx^{2}}=\frac{1}{\lambda_{J}^{2}}\sin\varphi
\end{gather}
and for the Abrikosov vortex situated at  $x_v=0$ and at the distance
$y_v$  from the junction  the Abrikosov vortex generates only the flux
$\Phi\left(  y_v\right)  $ and the remaining flux  ${\Phi_{0}-\Phi(
y_v) } $ is  generated by the Josephson vortex in the region
$\sim2\lambda_{J}$. Naturally the total flux of such a hybrid vortex is equal
to one quantum $\Phi_{0}$.

Magnetic field at the center of Josephson junction (we suppose that the thickness $d$ of the weak coupling region is much smaller than $\lambda$) is $h_z(x,y=0)=-\frac{\Phi_{0}}{4\pi\lambda}\nabla_x\varphi$ \cite{Barone}. Taking into account that it decays along the $y$-axis exponentially at distance $\lambda$ from the junction we get
\begin{gather}\label{h_JJ}
h_z(x,y)=\frac{\Phi_{0}}
{4\pi\lambda\lambda_{J}}\frac{e^{-|x|/\lambda_{J}}%
\tan\left(  \frac{\kappa\left(  y_v\right)  }{8}\right)  }{1+e^{-2\left\vert
x\right\vert /\lambda_{J}}\tan^{2}\left(  \frac{\kappa\left(  y_v\right)
}{8}\right)  }e^{-|y|/\lambda}.
\end{gather}
The characteristic magnetic field of such a partial Josephson junction $\sim\frac{\Phi_{0}}{\lambda\lambda_{J}}$ is much smaller than the field of Abrikosov vortex and can be hardly detected on experiment.
In Fig. \ref{fig3} we present the field distribution for such a composite Abrikosov[Eq. (\ref{h_long})] + Josephson[Eq. (\ref{h_JJ})] vortex for: (i) $y_v=0.7\lambda$ ($\kappa=\pi$), where half of the flux is carried by Abrikosov vortex and Josephson vortex; and (ii) $y_v=0.1\lambda$ ($\kappa=1.8\pi$), where almost all of the flux has been transferred to the Josephson vortex.

\textit{Vortex near the plane with $\mathcal{T}\approx1$}.
\begin{figure}[] 
\begin{minipage}[h]{1.0\linewidth}
\includegraphics[width=0.95\textwidth]{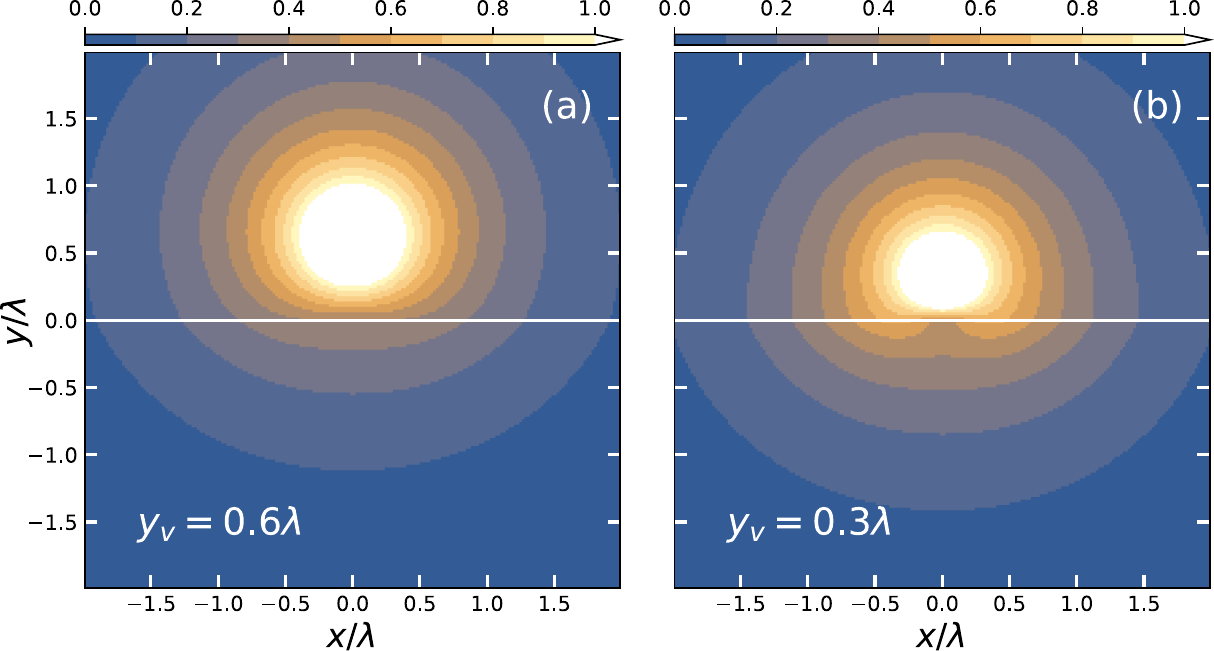} 
\end{minipage}
\caption{\small{ Distribution of the magnetic field $h_z/\frac{\Phi_0}{2\pi \lambda^2}$ of the Abrikosov vortex in the vicinity of the planar defect of the strength $\alpha=0.3$ (shown by the white line) for different vortex positions $y_v$.  }}
\label{fig4}
\end{figure}
As a last case we address the opposite limit of long (${\ell\gtrsim \lambda}$) planar defect of high transparency, which can be treated with the help of spatially modulated diffusion coefficient ${D({\bf r})\propto \lambda^{-2}({\bf r})}$ or London penetration depth. The local increase of the latter in the vicinity of the defect plane can be simply modeled as ${\lambda^2(y) = \lambda^2 + \alpha \lambda^2\delta(y)}$, where we introduced the dimensionless coefficient ${0 \leqslant  \alpha \propto (1-\mathcal{T})  \ll 1}$. Note that the energy of the vortex $U_p\propto \lambda^{-2}(y)$ decreases in the vicinity of the plane, therefore the local suppression of transparency ($\alpha>0$) leads to an attractive interaction. Local change of the supercurrent situated close to the defect plane ${\bf j} = (c/4\pi \lambda^2(y))( \frac{\Phi_0}{2\pi}\nabla \theta -{\bf A} )$
obeys the London equation which reads
\begin{gather}\label{LONDON}
    (\nabla^2-\lambda^{-2})h_z -\frac{\Phi_0}{2\pi\lambda^2}\delta({\bf r}_v)  \\ \notag 
    = \alpha\lambda\delta(y)\nabla^2h_z+\alpha\lambda\nabla_y\delta(y)\nabla_yh_z.
\end{gather}
Following approach of Ref. \cite{KoganMeissner} we treat the RHS of the Eq. (\ref{LONDON}) as a perturbation and obtain the distribution of the field $h_z({\bf r})=h_v({\bf r})+\alpha h_{z1}({\bf r})+\mathcal{O}(\alpha^2)$ using Fourier transform
\begin{gather}
h_z({\bf r}) = \frac{\Phi_0}{2\pi \lambda^2}K_0\Big[ \frac{|{\bf r}-{\bf r}_v|}{\lambda} \Big] \\ \notag
- \alpha\frac{\Phi_0}{2\pi \lambda^2}\frac{|y|+|y_v|}{\sqrt{(|y|+|y_v|)^2+x^2}} K_1\Big[ \frac{\sqrt{(|y|+|y_v|)^2+x^2}}{\lambda} \Big].
\end{gather}
%
%
%
Naturally, the field distribution is weakly extended along the defect with local anisotropy of the order of $\alpha$. It confirms the general tendency for the vortex magnetic field to spread with the decrease of $\mathcal{T}$, and this effect may be noticeable even in the intermediate regime $\mathcal{T}\lesssim 1$.

\textit{Discussion}.
We believe that the fractional vortices observed in Ref. \cite{FracVort} close to $T_c$ may be the Abrikosov vortices near the planar defects (which can appear in Ba$_{1-x}$K$_{x}$Fe$_{2}$As$_{2}$ due to the local doping variation). The remote field of such vortices may be too small for local magnetometry resolution above the noise level. Our results show that Abrikosov vortices are attracted to the planar defect and the attractive force (\ref{U_p}) increases with the decrease of the temperature through $\lambda(T)$, which may explain why the fractional vortices \cite{FracVort} disappeared below 9K  - they were absorbed by the defect or if the defect presents a local critical temperature increase, the short range repulsion will maintain vortex near the defect. The situation is somewhat similar to the vortex near the twinning plane with a local increase of superconducting pairing \cite{AbrBuzd}.

In conclusion, we have demonstrated that the planar defects of various transparency can strongly modify the vortex magnetic field. If the size of the defect exceeds the London penetration depth a substantial part of the magnetic flux of the vortex should be delocalized from its center which will affect the local ($\lesssim \lambda$) flux measurements.

We thank A. V. Samokhvalov and A. S. Mel'nikov for fruitful discussions. This work has been supported by ANR SUPERFAST, the LIGHT S$\&$T Graduate Program. A.B. acknowledges support by the Ministry of Science and Higher Education of the Russian Federation within the framework of state support for the creation and development of World-Class Research Center “Digital biodesign and personalized healthcare," No. 075-15-2022-304.

\end{document}